# Single-spacecraft techniques for shock parameters estimation: A systematic approach

D. Trotta[1]*, L. Vuorinen[2], H. Hietala[1,3], T. Horbury[1], N. Dresing[2], J. Gieseler[2], A. Kouloumvakos[4], D. Price[5], F. Valentini[6], E. Kilpua[5] and R. Vainio[2]

[1]The Blackett Laboratory, Department of Physics, Imperial College London, London, United Kingdom, [2]Department of Physics and Astronomy, University of Turku, Turku, Finland, [3]School of Physics and Astronomy, Queen Mary University of London, London, United Kingdom, [4]Applied Physics Laboratory, The Johns Hopkins University, Laurel, MD, United States, [5]Department of Physics, University of Helsinki, Helsinki, Finland, [6]Dipartimento di Fisica, Universita' della Calabria, Rende, Italy

*CORRESPONDENCE
D. Trotta,
d.trotta@imperial.ac.uk

Spacecraft missions provide the unique opportunity to study the properties of collisionless shocks utilising *in situ* measurements. In the past years, several diagnostics have been developed to address key shock parameters using time series of magnetic field (and plasma) data collected by a single spacecraft crossing a shock front. A critical aspect of such diagnostics is the averaging process involved in the evaluation of upstream/downstream quantities. In this work, we discuss several of these techniques, with a particular focus on the shock obliquity (defined as the angle between the upstream magnetic field and the shock normal vector) estimation. We introduce a systematic variation of the upstream/downstream averaging windows, yielding to an ensemble of shock parameters, which is a useful tool to address the robustness of their estimation. This approach is first tested with a synthetic shock dataset compliant with the Rankine-Hugoniot jump conditions for a shock, including the presence of noise and disturbances. We then employ self-consistent, hybrid kinetic shock simulations to apply the diagnostics to virtual spacecraft crossing the shock front at various stages of its evolution, highlighting the role of shock-induced fluctuations in the parameters' estimation. This approach has the strong advantage of retaining some important properties of collisionless shock (such as, for example, the shock front microstructure) while being able to set a known, nominal set of shock parameters. Finally, two recent observations of interplanetary shocks from the Solar Orbiter spacecraft are presented, to demonstrate the use of this systematic approach to real events of shock crossings. The approach is also tested on an interplanetary shock measured by the four spacecraft of the Magnetospheric Multiscale (MMS) mission. All the Python software developed and used for the diagnostics (SerPyShock) is made available for the public, including an example of parameter estimation for a shock wave recently observed *in-situ* by the Solar Orbiter spacecraft.

KEYWORDS

**space physics, shock waves, collisionless shocks, spacecraft data, heliosphere, python**

# 1 Introduction

The understanding of the processes controlling our Universe has always been intimately related to our capability to observe the large ensemble of phenomena taking place in astrophysical systems. Since the last century, our knowledge of the Universe has grown tremendously, thanks to the exceptional technological advances which made us able to observe it with great levels of detail. Such observational advances, together with important theoretical/modelling breakthroughs, reveal that some phenomena are universal, i.e., common to many astrophysical environments.

Among these universal phenomena, collisionless shock waves, i.e., abrupt transitions from super-magnetosonic (upstream) and sub-magnetosonic (downstream) flows, are particularly important to address various aspect of energy conversion in many astrophysical systems, ranging from solar flares (e.g., Benz, 2008) to the largest (Mpc) scales of galaxy cluster merging (see Brunetti and Jones, 2014, for a review).

In the late 1950s, thanks to the possibility of spacecraft flight in interplanetary space, the first collisionless shocks were observed in the heliosphere (see Kivelson and Russell, 1995, for a comprehensive introduction). The possibility to carry out *in-situ* shock observations has then become fundamental to address many of their properties and behaviour.

Shocks that are routinely observed in the heliosphere can be divided in several categories, namely interplanetary (IP) shocks, generated as a consequence of solar activity phenomena, such as Coronal Mass Ejections (CME) and Stream Interaction Regions (SIR) (Dessler and Fejer, 1963; Gosling et al., 1974), and planetary bow shocks, resulting from the interaction between the supersonic solar wind and planetary bodies behaving as obstacles for the solar wind flow (e.g., Dungey, 1979; Hoppe and Russell, 1982; Lepping, 1984). Other examples of shocks observed *in-situ* include the *Voyager* observations of the heliospheric termination shock at the interface between the heliosphere and the interstellar environment (e.g., McComas et al., 2019) and cometary bow shocks (Thomsen et al., 1986; Coates et al., 1997; Naeem et al., 2020). Despite the important differences between these different types of shocks, they share their fundamental, underlying physics.

The shock structure and behaviour is regulated by several parameters, the most important being the angle between the shock normal direction and the upstream magnetic field, $\theta_{Bn}$. When $\theta_{Bn}$ approaches 90°, the shock is quasi-perpendicular, i.e., the upstream magnetic field is along the shock front. On the other hand, for $\theta_{Bn}$ values close to 0° (corresponding to an upstream magnetic field almost normal to the shock surface), the shock is quasi-parallel. Particle reflection and propagation far upstream is favoured at quasi-parallel shocks (Kennel et al., 1985), introducing the possibility for reflected particles to interact with the upstream plasma over long distances, creating unstable distributions and a collection of disturbances in the plasma properties. Other important parameters for the shock behaviour are the shock Alfvénic and fast magnetosonic Mach numbers, i.e., the ratio between the shock speed in the upstream flow frame and the upstream Alfvén and fast magnetosonic speed, respectively ($M_A \equiv v_{sh}/v_A$ and $M_{fms} \equiv v_{sh}/v_{fms}$) (see Burgess and Scholer, 2015, for an extensive review).

Since the early *Pioneer* evidences of the Earth's bow shock existence (Dungey, 1979) to modern missions crossing various kind of heliospheric shocks (e.g., Masters et al., 2008; Blanco-Cano et al., 2016), various techniques to extract shock parameters as the ones described above from single spacecraft observations have been developed, starting from various theoretical formulations to describe the behaviour of the plasma properties across shock transitions (Lepping and Argentiero, 1971; Abraham-Shrauner, 1972; Viñas and Scudder, 1986). Further advancements on such investigations have been made using multiple satellite missions, such as *Cluster* and the Magnetospheric MultiScale (MMS) (Escoubet et al., 2001; Burch et al., 2016). Using multi-spacecraft measurements, important properties of shock transitions have been discovered (e.g., Johlander et al., 2016; Kajdič et al., 2019), and new parameter estimation techniques relying on multiple crossings have been developed (e.g., Russell et al., 1983a,b). However, many modern spacecraft missions, for example the recently launched Solar Orbiter mission, rely on a single spacecraft (Muller et al., 2020), making it important to understand the limits and advantages of shock parameters estimation techniques relying on the study of a single time series collected by the instrumentation on board of spacecraft during a shock crossing.

The understanding of collisionless plasmas in several environments, ranging from controlled fusion devices to astrophysical systems, has always been intimately related to the use of computer simulations, due to the complexity of the problems studied (e.g., Birdsall and Langdon, 1991). In fact, another crucial source of knowledge around collisionless shocks comes from simulation studies. Collisionless shocks have been modelled using several different methods, ranging from local and global magnetohydrodynamic (MHD) simulations studying the large scale properties of shock transitions (e.g., Mignone et al., 2007; Mejnertsen et al., 2018), to local simulations including kinetic effects, addressing the shock structuring and behaviour at small scales (e.g., Caprioli and Spitkovsky, 2014; Ha et al., 2022). Combining numerical and observational efforts, has often proved to be very effective for the understanding of shock behaviour (e.g., Sundberg et al., 2016).

As we shall see in the below discussion, most shock parameter estimation techniques from spacecraft data involve an operation of averaging plasma quantities upstream/downstream of the shock crossing, making the results particularly sensitive to the choice of averaging windows. The idea to adopt an ensemble-based approach for such choices of averaging windows was first suggested by Balogh et al. (1995), looking at magnetic field Ulysses observations of interplanetary shocks.

In this work, we review some of these techniques, proposing a systematic way of varying the averaging windows used for such parameter estimations. The importance and implications of this approach are discussed, and the approach is then tested on a model shock. We then propose the important, novel test of these parameter estimation techniques on a shock simulated using state-of-the-art, self-consistent plasma simulations, in which we have the opportunity to investigate the shock geometry at different stages of its evolution. Finally, two different real world observations are presented, obtained in late 2021 by the Solar Orbiter spacecraft, and two applications of the systematic shock parameters estimation are shown.

Recently, the importance to share tools for data analysis is object of discussion across several scientific communities. A virtuous example of such an effort in the space physics community is the publicly available software for spacecraft data analysis irfu-matlab (https://github.com/irfu/irfu-matlab). We make the Python software developed in this work and used for the systematic approach to *in-situ* shock analyses (SerPyShock) publicly available at https://github.com/trottadom/SerPyShock. This code release includes the routines for the analysis, an interactive test case for the Rankine-Hugoniot compliant shocks shown below, and an example of analysis on a Solar Orbiter dataset, for which Python data loaders are also provided. The software is written using the PEP8 standard, in Python 3.9 and requires a minimal number of standard dependencies, e.g., numpy and scipy, listed in the repository. Future updates to the software will include further toold of data analysis, such as for example an algorithm for particle/wave foreshock identification.

The paper is organised as follows: a brief summary of the theoretical framework is presented in Section 2; in Section 3, the main results of this work are shown, discussing the analytical, simulation and observational cases in Sections 3.1–3.3, respectively; the conclusions are in Section 4. In this work, particular emphasis is given to addressing the shock obliquity and in particular the computation of the $\theta_{Bn}$ angle. Software to compute other parameters is also provided and briefly discussed in the Appendix.

## 2 Theoretical framework

Below, we briefly review the theoretical framework underlying the parameter estimation techniques object of this work. To do so, we closely follow Chapter 10 of Paschmann and Schwartz (2000), Burgess and Scholer (2015) and Hietala (2012).

This discussion starts from a widely used approach to model the plasma properties across a shock, namely the Rankine-Hugoniot jump conditions (Rankine, 1870; Hugoniot, 1887). An MHD model is built for the shock, treated as an abrupt transition between the convected upstream plasma (supersonic) to downstream (subsonic). Here, the term "abrupt" may seem slightly vague, but is inspired by the consideration that the thickness of the transition must be related to a typical dissipation scale, that is much smaller than the typical convection scale in high temperature plasmas.

Let us consider the shock as a discontinuity and move into the shock rest frame (i.e., a reference frame in which the shock transition has speed equal to zero). In this frame, two fundamental assumptions are made, namely the fact that the shock front is spatially uniform, and time stationary. If we imagine the transition from supersonic to subsonic flow happening along the $x$ direction of a cartesian system, the above assumptions imply that $\partial/\partial t = \partial/\partial y = \partial/\partial z = 0$. Furthermore, due to the above, the shock thickness is considered to be infinitesimal in this model. At this point, the MHD equation for mass, momentum and energy conservation, and the divergence-free condition for the magnetic field are considered and integrated, yielding to the following relations:

$$[\rho_m V_n] = 0 \tag{1}$$

$$\left[\rho_m V_n^2 + P + \frac{B_t^2}{2\mu_0}\right] = 0 \tag{2}$$

$$\left[\rho_m \mathbf{V}_t V_n + P + \frac{\mathbf{B}_t B_n}{\mu_0}\right] = 0 \tag{3}$$

$$\left[\frac{1}{2}\rho_m V^2 V_n + \frac{\gamma}{\gamma - 1} P V_n + \frac{B_t^2}{\mu_0} V_n - \frac{\mathbf{V}_n \cdot \mathbf{B}_t}{\mu_0} B_n\right] = 0 \tag{4}$$

$$[B_n] = 0 \tag{5}$$

$$[\mathbf{V}_t B_n - \mathbf{B}_t V_n] = 0. \tag{6}$$

Where $\rho_m$ is the mass density, $\mathbf{V}$ is the bulk flow speed, $P$ is the scalar thermal pressure of the plasma, $\mathbf{B}$ the magnetic field. $\gamma$ and $\mu_0$ are the polytropic index and the vacuum permeability, respectively. Here, the subscripts $n$ and $t$ indicate that the quantity is considered normal to the shock and transverse to it, respectively. In the equations above, the symbols $[\ldots]$ indicate the difference between downstream and upstream quantities, e.g., $[F] = F_2 - F_1$ for a generic field $F$, where the subscript 2 and 1 indicate that the field is evaluated downstream and upstream, respectively. The Rankine-Hugoniot relations 1-6 are of crucial importance to address shock properties in a variety of systems.

We now consider the shock geometry to be oblique (i.e., $\theta_{Bn} \neq 90°, 0°$), and move in de Hoffmann-Teller frame, i.e., a frame aligned to the shock normal that moves at a speed such that the upstream convective electric field ($\mathbf{E}_1 = -\mathbf{V}_1 \times \mathbf{B}_1$) vanishes (de Hoffmann and Teller, 1950). Such a frame is extremely important for shock physics, especially to understand various properties of accelerated particles across shock transitions (e.g., Leroy and Mangeney, 1984).

Unrolling the [...] operation in Eqs 1–6 for an oblique shock in the de Hoffmann-Teller frame, the Rankine-Hugoniot jump conditions become:

$$\frac{\rho_{m2}}{\rho_{m1}} = \mathrm{r_{gas}} \tag{7}$$

$$\frac{V_{n2}}{V_{n1}} = \frac{1}{\mathrm{r_{gas}}} \tag{8}$$

$$\frac{V_{t2}}{V_{t1}} = \frac{M_{A1}^2 - 1}{M_{A1}^2 - \mathrm{r_{gas}}} \tag{9}$$

$$\frac{B_{n2}}{B_{n1}} = 1 \tag{10}$$

$$\frac{B_{t2}}{B_{t1}} = r\frac{M_{A1}^2 - 1}{M_{A1}^2 - \mathrm{r_{gas}}} \tag{11}$$

$$\frac{P_2}{P_1} = \mathrm{r_{gas}} + \frac{(\gamma-1)\mathrm{r_{gas}}V_1^2}{2V_S1}\left(1 - \frac{V_2^2}{V_1^2}\right). \tag{12}$$

Here, two important shock parameters arise: $\mathrm{r_{gas}}$, i.e., the shock gas compression ratio (notation has been chosen to distinguish it from the shock magnetic compression ratio $\mathrm{r_B} \equiv \frac{B_2}{B_1}$) and $M_{A1} \equiv V_{n1}\sqrt{\mu_0\rho_{m1}}/B_{n1}$. $V_{S1}$ is the upstream sound speed. As it is known, several sound speeds are admitted in the MHD formulation for wave propagation, and indeed the MHD system admits three different types of shocks, namely slow, fast and intermediate. We focus here on fast shocks, and refer the reader to Burgess and Scholer (2015) for a detailed discussion.

At this point, the Rankine-Hugoniot conditions Eqs 7–12 can be used to address the shock geometry. This is done in conjunction with the use of the coplanarity theorem, stating that the upstream, downstream magnetic field and shock normal vectors lie all in the same plane (see Eqs 10–11). It is with such considerations in mind that we introduce the first diagnostic for computing the shock normal, namely the magnetic coplanarity (MC) method (Colburn and Sonett, 1966):

$$\hat{n}_{\mathrm{MC}} = \pm\frac{(\mathbf{B}_2 \times \mathbf{B}_1) \times \Delta\mathbf{B}}{|(\mathbf{B}_2 \times \mathbf{B}_1) \times \Delta\mathbf{B}|}. \tag{13}$$

As it can be seen, this diagnostic relies on upstream/downstream magnetic field estimation only, with $\Delta\mathbf{B} = \mathbf{B}_2 - \mathbf{B}_1$. The shock normal sign obtained with such a method is arbitrary, and it is adjusted by convention such that $\hat{n}$ points upstream, hence the ± sign. Through a corollary of the coplanarity theorem, it is possible to see that the vector $\Delta\mathbf{V}$, i.e., the difference between downstream and upstream velocity lies also in the same plane as $\mathbf{B}_2$, $\mathbf{B}_1$ and $\hat{n}$ (see Paschmann and Schwartz, 2000). Using this corollary, it is possible to introduce the following additional methods for the shock normal estimation, namely the three *mixed mode* (MX1, MX2, MX3) methods. These combine magnetic field and bulk flow speed measurements across the shock transition:

$$\hat{n}_{\mathrm{MX1}} = \pm\frac{(\mathbf{B}_1 \times \Delta\mathbf{V}^{\mathrm{arb}}) \times \Delta\mathbf{B}}{|(\mathbf{B}_1 \times \Delta\mathbf{V}^{\mathrm{arb}}) \times \Delta\mathbf{B}|} \tag{14}$$

$$\hat{n}_{\mathrm{MX2}} = \pm\frac{(\mathbf{B}_2 \times \Delta\mathbf{V}^{\mathrm{arb}}) \times \Delta\mathbf{B}}{|(\mathbf{B}_2 \times \Delta\mathbf{V}^{\mathrm{arb}}) \times \Delta\mathbf{B}|} \tag{15}$$

$$\hat{n}_{\mathrm{MX3}} = \pm\frac{(\Delta\mathbf{B} \times \Delta\mathbf{V}^{\mathrm{arb}}) \times \Delta\mathbf{B}}{|(\Delta\mathbf{B} \times \Delta\mathbf{V}^{\mathrm{arb}}) \times \Delta\mathbf{B}|}, \tag{16}$$

Where $\Delta\mathbf{V}^{\mathrm{arb}} \equiv \mathbf{V}_2 - \mathbf{V}_1$, and the arb superscript indicates that the velocity difference is calculated in an arbitrary reference frame. Mixed mode methods have been extremely popular in addressing shock geometries in previous literature (e.g., Volkmer and Neubauer, 1985; Balogh et al., 1995; Kilpua et al., 2015), including a recent effort looking at geomagnetic activity triggered by shocks with different inclinations (Oliveira and Samsonov, 2018). There are also other methods, which are not discussed in this work, to estimate the shock normal. These are, for example, the velocity coplanarity method (Abraham-Shrauner, 1972) and the minimum variance analysis (Sonnerup et al., 1967). Another class of diagnostics not discussed here are the ones dealing with multi-spacecraft observations (Paschmann and Schwartz, 2000).

Even though we focus mostly on techniques estimating shock geometry, an algorithm for shock speed estimation is here reviewed (and included in the SerPyShock software released within this work), namely the mass flux algorithm, defined as:

$$V_{\mathrm{sh}}^{\mathrm{arb}} = \frac{\Delta(\rho\mathbf{V}^{arb})}{\Delta\rho}\cdot\hat{n}. \tag{17}$$

The definition above, obtained applying the mass flux conservation equation across the shock discontinuity, yields to an estimation of the shock speed along the shock normal. As above, the Δ symbol indicates that a difference between the downstream and the upstream quantity is performed. An important limitation of the mass flux algorithm is the fact that is reliant on plasma measurements only. An example of shock speed estimation using the mass flux algorithm, systematically varying the averaging windows, as discussed in detail in this work in the context of shock geometry estimation, is shown in the Appendix of this manuscript. Finally, we note that an interesting approach to parameter estimation, using the so-called modified Rankine-Hugoniot shock fitting technique, was proposed by Viñas and Scudder (1986) and since applied successfully to spacecraft data (Koval and Szabo, 2008), yielding to a simultaneous estimation of shock normal and speed.

## 3 Methodology and results

Spacecraft measured quantities consist of time-series data streams. As discussed above, all the methodologies discussed so far rely on a crucial choice, namely the extent of upstream/downstream averaging windows for the quantities involved in the diagnostics. Clearly, the parameters deducted from the observations will depend on the averaging windows sizes, and the parameter estimation is robust when the results do not change for different window choices. Such approach is here formalised and carried out systematically, then tested on different theoretical and observational examples.

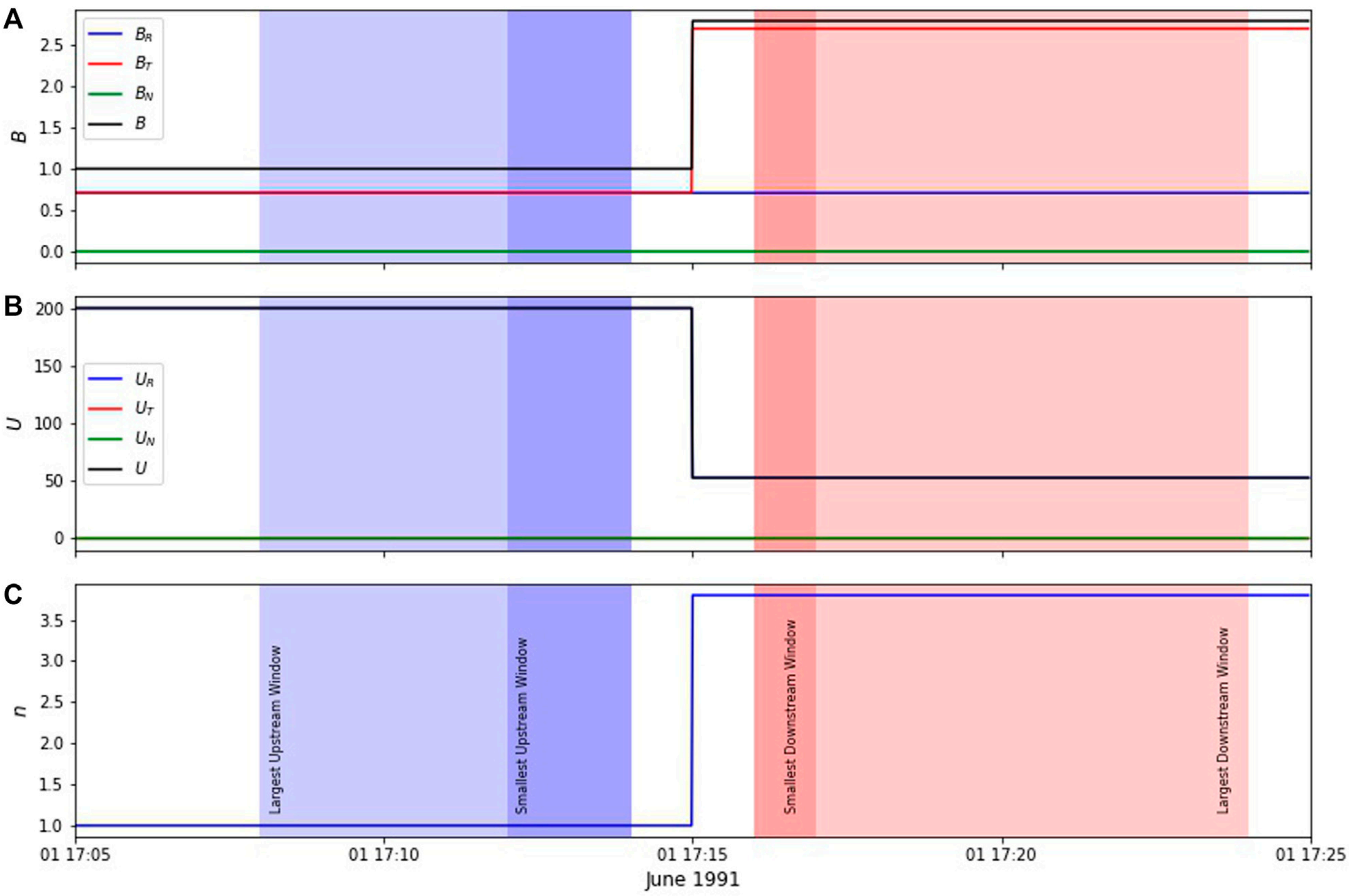

FIGURE 1
Synthetic measurements for a Rankine-Hugoniot compliant shock with $\theta_{Bn}$ = 45° and gas compression ratio $r_{gas}$ = 3.8. Panel show magnetic field magnitude and components **(A)**, plasma bulk flow speed **(B)** and density **(C)**. The blue (red) shaded areas highlight the windows chosen for upstream (downstream) averaging.

## 3.1 Rankine-Hugoniot compliant synthetic field

For the introduction and first testing of our approach, a synthetic spacecraft measurement was generated using the Rankine-Hugoniot jump conditions discussed in Section 2.

Results are shown in Figure 1, where such synthetic measurements for a Rankine-Hugoniot compliant shock with $\theta_{Bn}$ = 45° and gas compression ratio $r_{gas}$ = 3.8 is represented. As it can be seen from magnetic field, bulk flow speed and density time-series measurements (top, middle and bottom of Figure 1, respectively), the shock transition is such that the upstream is on the left-hand side of the Figure, i.e. before 17:15 of 1 June 1991, and the downstream is represented by the timeseries after 17:15 (obviously, time here has no physical meaning).

The parameter estimation is performed as follows: a smallest possible averaging window is chosen for upstream and downstream. Given the discussion about the Rankine-Hugoniot relations assuming the shock transition as infinitesimal, the idea is to choose such windows as close as possible to the shock front, without including it in the averaging process. Care must be taken in excluding also the shock foot in the very close upstream, as well as the downstream overshoot. When applying these diagnostics, as it will be discussed later concerning real observations, the choice for smallest/largest upstream window will depend on many different factors, e.g., the amount of disturbances that may be present upstream/downstream of the shock and the resolution available for the measurements. Another advice is that it is preferable to choose the smallest (largest) averaging windows to be of comparable scale sizes, thus avoiding systematic mixing of very different scales.

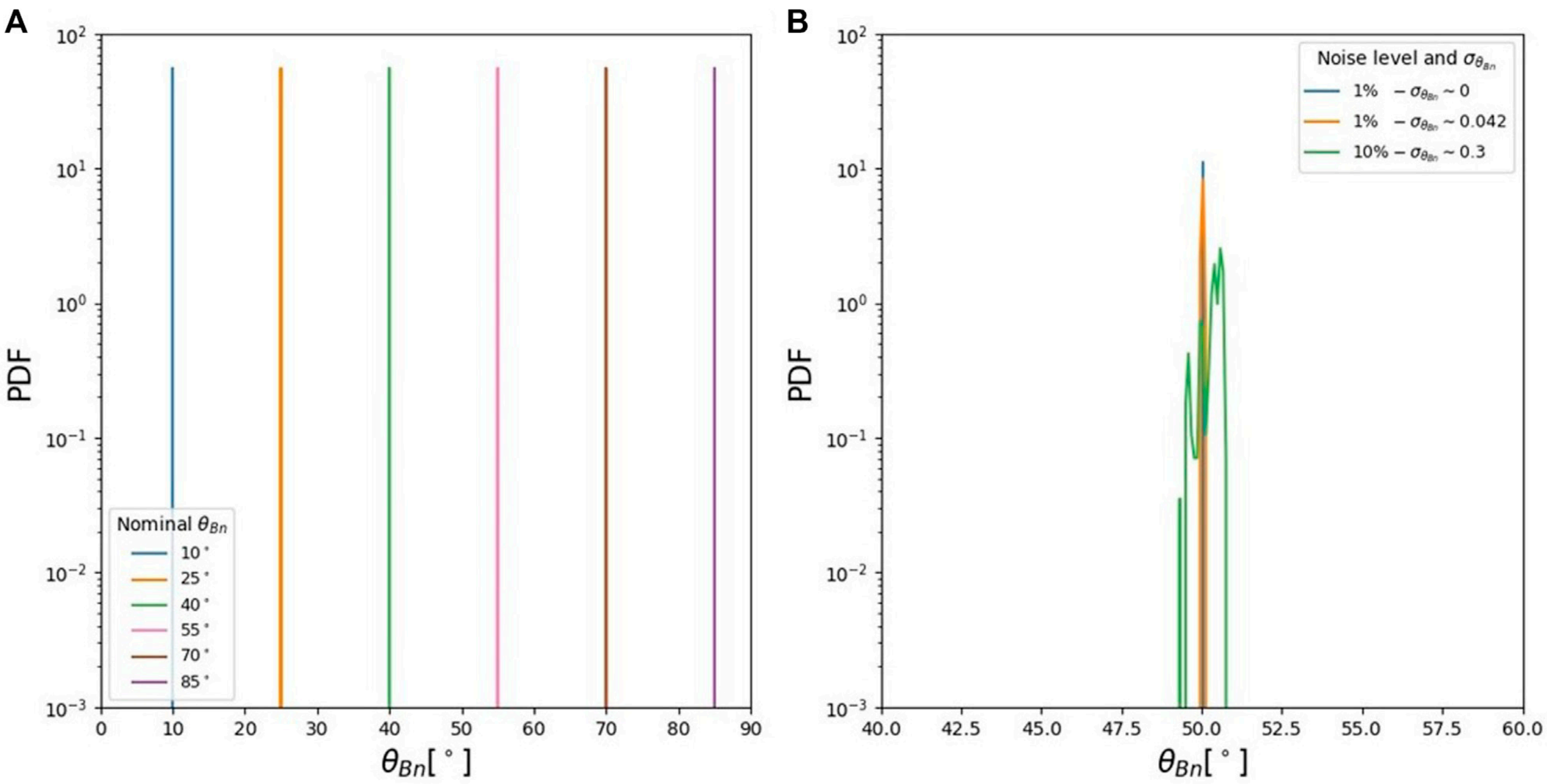

FIGURE 2
**(A)** PDF distributions for $\theta_{Bn}$ computed using magnetic coplanarity and mixed mode techniques, for six different synthetic, Rankine-Hugoniot compliant shocks. **(B)** PDF distributions of $\theta_{Bn}$ for three shocks with the same nominal geometry and different levels of added noise.

In Figure 1, the smallest upstream (downstream) averaging windows are shown with the dark shaded blue (red) panels. Then, after choosing an appropriate cadence, the smallest averaging windows are extended to the largest, and overlapped with the smallest windows. For each couple of upstream/downstream windows, a shock normal (and subsequently $\theta_{Bn}$ value) is evaluated. In this way, an ensemble of shock parameters is computed, and it is possible to address the robustness of the parameters estimation by looking at how is the ensemble distributed, as we shall see below. Note that, since we add duration to the smallest upstream/downstream windows, the measurements close to the shock are counted multiple times, ensuring that the calculation is carried out as close as possible to the shock front at all times.

A first example of such a systematic approach is shown below. We considered 6 different Rankine-Hugoniot compliant shocks with different $\theta_{Bn}$ (10, 25, 40, 55, 70 and 85°, respectively), with the same approach described above. A collection of about 1,000 different windows have then been used to compute the shock geometry.

The results of this experiment have been reported in Figure 2A, where the probability density functions (PDFs) for the computed $\theta_{Bn}$ values are shown. Such PDFs, as expected, are peaked (within machine precision) at the nominal $\theta_{Bn}$ value for each shock. This test is useful for purposes of routine testing, as well as for a sanity check of the code that is then used for real-world cases. Furthermore, this case and the code used to generate the Ranking-Hugoniot compliant shocks is released as an interactive test for the SerPyShock user.

At this stage, the robustness of the approach has been tested on slightly more complicated case. Three synthetic shocks have been generated, all of them with the same $\theta_{Bn} = 50^\circ$, considering the baseline case of Figure 2A with other two cases where white noise of 1% and 10% level was added, respectively (Figure 2B). In this case, it can be seen that when noise is added, the PDF distribution of $\theta_{Bn}$ values broadens, as it is particularly evident for the 10% noise level case (green line in Figure 2B). Another interesting feature of this method is that it is possible to characterise the spread of this PDFs computing, together with the average value $\langle\theta_{Bn}\rangle$, its standard deviation $\sigma_{\theta_{Bn}}$, which becomes a measure for the sensitivity of the parameter estimation to the choice of upstream/downstream window and, in other words, an uncertainty for the parameter estimation. It is possible to note that the broadening observed here is rather small (with a maximum discrepancy of about 1° from the nominal shock geometry), due to the fact that the averaging process averages out the white noise. Furthermore, it is possible to chack if the broadening of the distribution follows a gaussian profile by computing the skewness of the PDFs of shock parameters. In the added noise case of these synthetic shocks, a very small values of skewness is always found, with the Fisher-Pearson coefficients of skewness (see Kokoska and Zwillinger, 1999) being $g_1 = 0.0, 2 \times 10^{-3}$ and $3 \times 10^{-2}$ for the 0%, 1% and 10% noise cases, respectively.

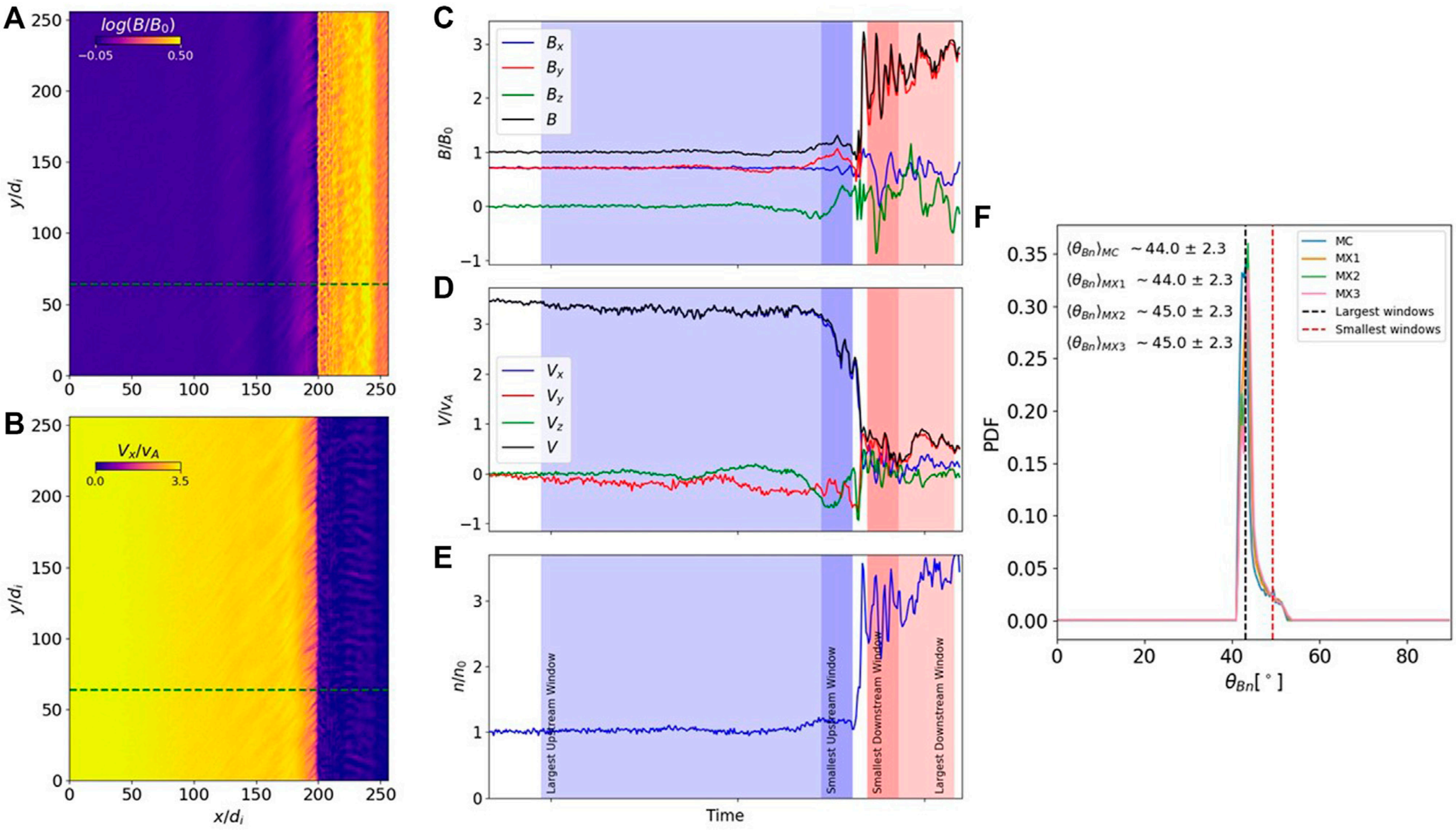

**FIGURE 3**
Snapshot of Hybrid PIC simulation of 45° degrees shock captured at simulation time $T\Omega_{ci} = 40$. Panels **(A)** and **(B)** show magnetic field magnitude and proton bulk flow speed along the shock nominal shock normal direction, respectively. In the middle panels, are shown one-dimensional signals of magnetic field and its components **(C)**, proton bulk flow speed magnitude and its components **(D)** and proton density **(E)** as seen by a virtual spacecraft crossing the shock along the green-shaded line in panels **(A,B)**. In these panels, the smallest/largest averaging windows used for the upstream/downstream regions are highlighted (red/blue shaded panels). Finally, PDF distributions of $\theta_{Bn}$ obtained with MC, MX1, MX2 and MX3 methods are shown in **(F)**.

## 3.2 Hybrid PIC simulations

Here, we test the methods to compute the shock obliquity with the systematic window variation approach on a self-consistent, kinetic simulation of a supercritical shock. For this experiment, we use the HYPSI code, successfully used in the past for several studies addressing collisionless shocks (e.g., Trotta and Burgess, 2019; Preisser et al., 2020a; Trotta et al., 2021).

In the simulation, protons are modelled as macroparticles and advanced using the standard PIC method. The electrons, on the other hand, are modelled as a massless, charge-neutralizing fluid with an adiabatic equation of state. The HYPSI code is based on the Current Advance Method and Cyclic Leapfrog (CAM-CL) algorithm (Matthews, 1994). The shock is initiated by the injection method (Quest, 1985), in which the plasma flows in the $x$-direction with a defined (super-Alfvénic) velocity $V_{\rm in}$. The right-hand boundary of the simulation domain acts as a reflecting wall, and at the left-hand boundary plasma is continuously injected. The simulation is periodic in the $y$ direction. A shock is created as a consequence of reflection at the wall, and it propagates in the negative $x$-direction. In the simulation frame, the (mean) upstream flow is along the shock normal.

In the hybrid simulations, distances are normalised to the ion inertial length $d_i \equiv c/\omega_{pi}$, times to the inverse cyclotron frequency $\Omega_{ci}{}^{-1}$, velocity to the Alfvén speed $v_A$ (all referred to the upstream state), and the magnetic field and density to their upstream values, $B_0$ and $n_0$, respectively. For the upstream flow velocity, a value of $V_{\rm in} = 3.5v_A$ has been chosen, and the resulting Alfvénic Mach number of the shock is approximately $M_A \sim 5$. The upstream ion distribution function is an isotropic Maxwellian and the ion $\beta_i$ is 1. The simulation $x - y$ domain is $256 \times 256$ $d_i$, and the simulation is therefore 2.5-dimensional. The spatial resolution used is $\Delta x = \Delta y = 0.5$ $d_i$. The system is evolved for 70 $\Omega_{ci}^{-1}$, with a time step for particle (ion) advance of $\Delta t = 0.01$ $\Omega_{ci}^{-1}$. Substepping is used for the magnetic field advance, with an effective time step of $\Delta t_B = \Delta t/10$. A small, nonzero resistivity is introduced in the magnetic induction equation, with a value that is chosen such that there are not excessive fluctuations at the grid scale. The number of particles per cell used is always greater than 500 (upstream), in order to keep the statistical noise characteristic of PIC simulations to a reasonable level. For the simulation studied here, the upstream magnetic field is in the $x$-$y$ plane, resulting in a nominal $\theta_{Bn}$ of the shock of 45°. As we shall see below, this is an excellent test case for our method, but analogous results can be obtained for different nominal shock geometries.

Figure 3 shows an overview of the shock simulation, at time $T\Omega_{ci}$ = 40. Panels (a) and (b) highlight the main features of the shock transition, showing magnetic field magnitude and proton bulk flow speed along the $\hat{x}$-direction, i.e., the shock normal nominal direction. Here, it can be seen that the shock front appears marginally stable, with small scale rippling due to its supercritical nature (e.g., Johlander et al., 2016). The presence of reflected particles streaming away from the shock front in the upstream region induces the upstream fluctuations in the vicinity of the shock. The shock transition is then shown through the one-dimensional plots in panels (c)-(e), that represent horizontal slices through the simulation domain along the green lines in panels (a) and (b). Such one-dimensional signals are the ones that a (virtual) spacecraft would observe while crossing the shock transition, provided that the spatial information can be translated into time information through the so-called Taylor hypothesis (Taylor, 1938), widely used in the context of space plasma observations (see Perri et al., 2017, for a detailed study of its validity).

At this point, the shock obliquity is calculated with the method described above, for the one-dimensional simulated spacecraft trajectory. The smallest and largest windows used for the upstream/downstream averaging are displayed using the shaded areas of Figure 3C–E. The result of the analysis is displayed in Figure 3F, where the PDF distributions of the $\theta_{Bn}$ values obtained with MC and MX1-2-3 techniques are shown. The errors shown together with the $\langle\theta_{Bn}\rangle$ in Figure 3F correspond to the standard deviation $\sigma_{\theta_{Bn}}$ for each estimation. Here, in contrast with the analytical examples shown above, a wider distribution of $\theta_{Bn}$ values is observed. Importantly, the PDF distributions yield to an average $\theta_{Bn}$ value of 45°, as set up in the simulation initialisation. In this case, the MC and MX methods agree very well. The spread of the distribution extending to higher values of $\theta_{Bn}$ is then due to the field oscillations present in the upstream/downstream shock regions. We would like to underline the importance of this example for the problem addressed in this work, from two points of view. First of all, it is demonstrated in numerical simulations, that the shock environment is often much more complicated with respect to the assumptions that go into the use of shock geometry estimation. Secondly, it is shown that to use the ensemble approach for the averaging windows choice is an optimal choice, as the result converge to the nominal shock geometry. Contrarily to this systematic approach, choosing two windows would only get one value for the $\theta_{Bn}$ measurement, and this may depart significantly to the average shock geometry. Such departures, due to shock front instabilities and various field-particle interaction, are interesting in their own respect, studied in recent literature (e.g., Kajdič et al., 2019; Preisser et al., 2020b).

From Figure 3, it is intuitive to link the spread of obtained $\theta_{Bn}$ values to the level of upstream/downstream fluctuations and disturbances. It is then natural to investigate how results change throughout different stages of the shock evolution. This is done in Figure 4, where the parameter estimation has been carried out at different simulation times. As the shock travels in the negative $x$-direction of the simulation domain, it is possible to see several microinstabilities being excited along the shock front. The presence of reflected particles, generating unstable upstream particle distributions, induce upstream fluctuations, more and more evident for later times, that are then convected downstream, generating a complex scenario for the shock transition.

When shock obliquity is addressed at the different shock evolution stages, the PDF distribution of the observed $\theta_{Bn}$ value broadens significantly. At $T\Omega_{ci}$ = 20, with a quasi-laminar shock transition (Figure 4D), the PDF is strongly peaked at the nominal value of $\theta_{Bn}$ = 45°, with a tail departing from it probably due to microinstabilities very close to the shock front. The scenario changes dramatically for the well-developed shock transition case $T\Omega_{ci}$ = 70, where the PDF is spread across a very large range of values. The skewness coefficients for the three crossings are $g_1 = -1.5, -1.3, -0.43$, indicating that the asymmetry due to small scale instabilties is stronger at early times and decreases at $T\Omega_{ci}$ = 70, where a larger spread of values is observed. Very importantly, the PDF distributions average very close to the nominal $\theta_{Bn}$ value, while the $\sigma_{\theta_{Bn}}$ increases for increasingly disturbed scenarios. This feature highlights the parameter estimation capabilities provided by the systematic variation of averaging windows.

To further investigate the dependence of the parameter distributions on the shock crossing performed by the virtual spacecraft, the inset of Figure 4E shows the total PDF of $\theta_{Bn}$ (black line) for the virtual spacecraft crossing in Figure 4B, together with the $\theta_{Bn}$ PDF obtained using only very small windows (max ~ $3d_i$, red line). It is therefore clear that the bump in the distribution is due to the virtual spacecraft crossing the shock along a ripple that is rotated towards the upstream field, i.e., a quasi-parallel portion of the shock front. This information is then lost when averaging over larger scales. This analysis shows that the systematic variation of averaging windows may be leveraged to obtained further information about shock structuring. Future upgrades of our software will include the possibility to cross-correlate estimated parameters and upstream/downstream window length. However, it is worth underlining that the spacecraft observations provide a more complex scenario, in which the multi-dimensional picture of the shock is not available, and the presence of pre-existing fluctuations can be hard to disentangle from shock-produced features of the upstream/downstream medium. Thus, care must be taken when interpreting such distributions of parameters, as it will be discussed in Sections 3.3, 3.4.

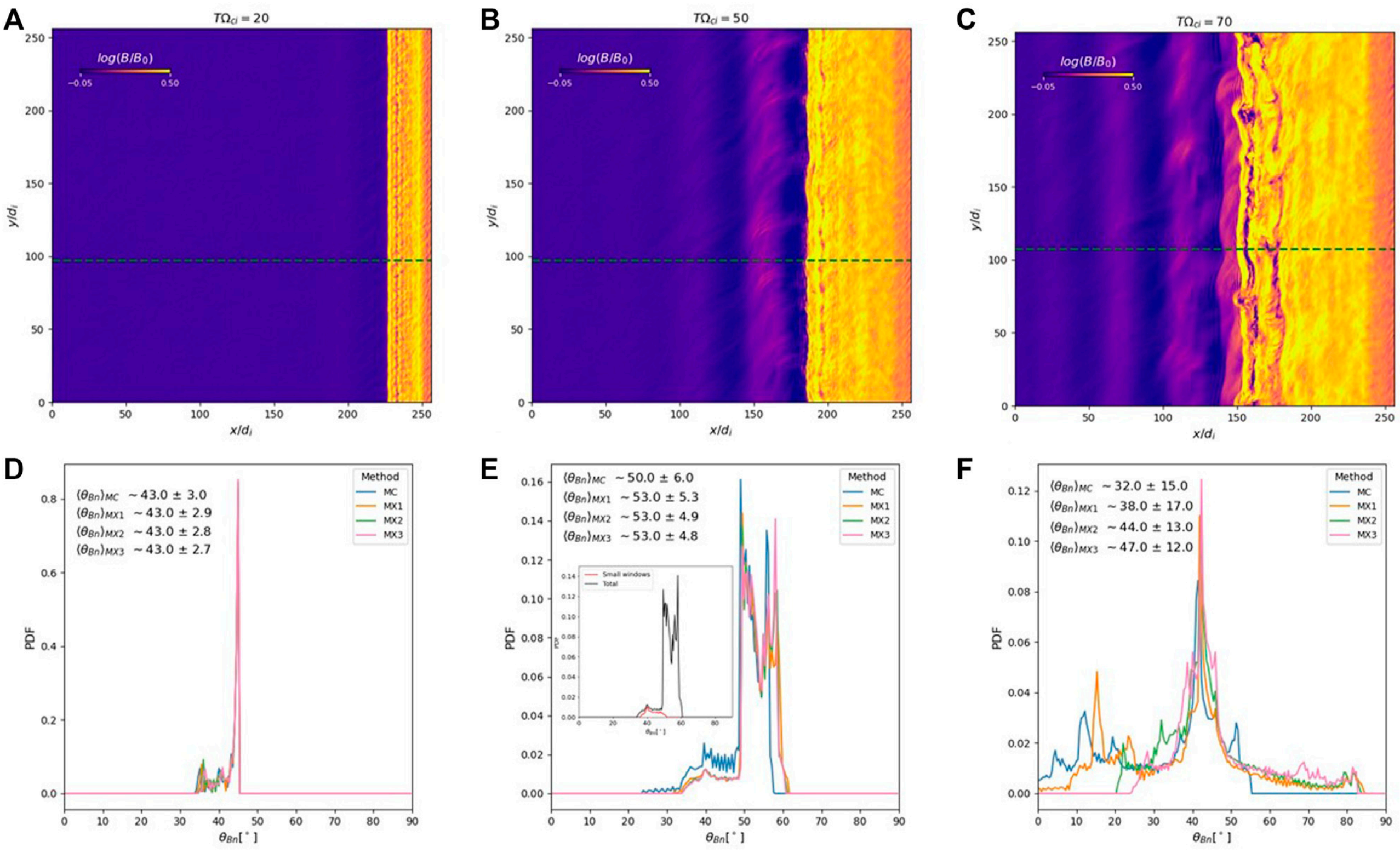

FIGURE 4
Magnetic field magnitude at different stages of shock evolution **(A–C)** and PDF distributions of $\theta_{Bn}$ values **(D–F)** evaluated on virtual spacecraft signals (with trajectories along green dashed lines) using systematic averaging window variations. The inset in panel **(E)** shows the full distribution of $\theta_{Bn}$ values (black) together with the distribution of $\theta_{Bn}$ obtained with small windows (max 2 $d_i$) only.

## 3.3 Solar orbiter shock observations

In the previous section, the diagnostics have been tested on a virtual spacecraft signal obtained using a self consistent simulation, which locally reconstructs the properties of a supercritical shock transition. Even though such simulations have been successful at improving our understanding of many observational features of shock transitions, they still represent an idealisation of very small spatial and temporal behaviour for the plasma. In this section, we finally apply our systematic method for shock parameter estimation on two events recently observed by the Solar Orbiter spacecraft.

Two different instruments on-board of Solar Orbiter have been used. The magnetic field, measured with a resolution of 128 vectors/s in burst mode by the flux-gate magnetometer MAG (Horbury et al., 2020), and the ground computed plasma moments, namely ion bulk flow, density and temperature, measured by the Solar Wind Analyser (SWA) suite (Owen et al., 2020), with 4 s resolution.

The first shock analysed here is a Coronal Mass Ejection (CME)-driven shock that crossed Solar Orbiter, while it was at 0.69 AU from the Sun at 07:32 UT of 11 October 2021. This shock is characterised by rather low Alfvén and fast magnetosonic Mach numbers ($M_{fms}$ ~ 2.04, $M_A$ ~ 2.5), low gas compression ratio $r_{gas}$ ~ 1.74 and a low level of upstream magnetic fluctuations $\langle \delta B/B_0 \rangle_{rms} \sim 0.08$, where $B_0$ indicates the magnetic field averaged upstream. Inspired by such a parameter set, indicating a subcritical shock transition, together with the absence of strong pre-existing Solar Wind structures surrounding the shock, this event was named the “quiet” event.

Figure 5 shows an overview for the quiet event, with ~ 30 min of spacecraft collected data. The shock transition, highlighted by the vertical magenta line in the Figure, appears well-behaved, without strong upstream/downstream structuring. In particular, low levels of upstream/downstream fluctuations make this a good observational case for the systematic shock parameters estimation method.

The parameter estimation reveals that the shock geometry is quasi-perpendicular, with $\langle \theta_{Bn} \rangle \sim 60°$. This has been obtained using smallest averaging windows of about 2 min upstream and downstream, and largest windows of about 10 min. This choice is such that the time window over which the average is taken is always larger than kinetic timescales ($\Omega_{ci}^{-1}$ is of order ~ 10 s), making sure that the MHD description on which the data analyis technique rely is appropriate. Other studies looking at IP shock statistics and catalogues have analogous choices for upstream/downstream windows (see, for example, http://www.ipshocks.fi). The upstream/downstream averaging windows have been broadened with timesteps of 16 s, larger than the resolution of the Solar Orbiter plasma instrument. As it can be seen from the right-hand side panel of Figure 5, the PDF distribution of $\theta_{Bn}$ values is strongly peaked, with a small value of $\sigma_{\theta_{Bn}}$ ~ 2. The skewness of the distribution is about 0.7, due to the secondary peak observable in the distribution at for $\theta_{Bn}$ around 68°. Furthermore, we find strong agreement between the results obtained using different techniques MC and MX1-2-3 for the shock normal evaluation. These results indicate that the parameter estimation for this shock transition is particularly robust, i.e., it has weak dependence on the choice of upstream/downstream averaging windows.

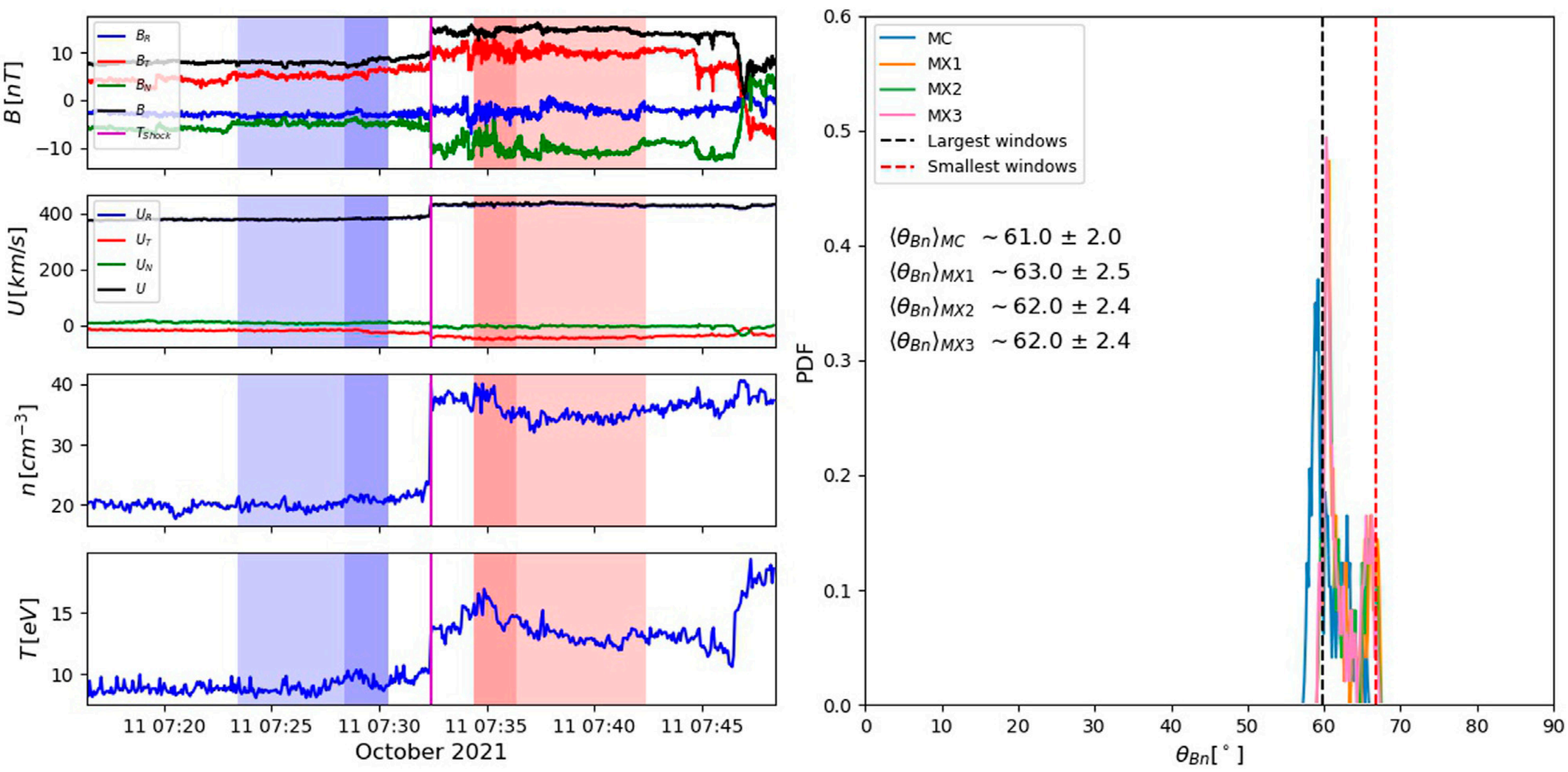

FIGURE 5
The quiet event of 11 October 2021 seen by Solar Orbiter. The left-hand panels show magnetic field $B$, ion bulk flow $U$, ion density $n$ and temperature $T$ measured by the MAG and SWA instruments. Vector quantities are shown in the RTN coordinate system. The blue/red shaded panels show the smallest (dark) to largest (lighter) averaging windows in the upstream/downstream regions. The vertical purple line shows the shock crossing time. The right-hand side panel shows the PDF distribution of $\theta_{Bn}$ values obtained for this event, with the vertical red (blue) dashed line showing the parameter esimation using the smallest (largest) window choice. The average values of $\theta_{Bn}$ obtained with each technique are also shown.

A different situation is observed for the second shock crossing event presented here. This is another CME-driven shock, that crossed Solar Orbiter a few days later than the quiet event, at 22:02 on 30 October 2021, while Solar Orbiter was at 0.82 AU from the Sun. This shock is stronger than the previous one, with $M_{fms}$ ~ 4.66 and $M_A$ ~ 6.92. For the gas compression ratio, we found $r_{gas}$ ~ 2.68, and the level of upstream magnetic fluctuations is moderate $\langle \delta B/B_0 \rangle_{rms}$ ~ 0.2.

The ~30 min' event overview, shown in Figure 6, shows a much more structured shock crossing, characterised by high levels of fluctuations downstream, as well as a rather disturbed upstream region. This behaviour has a strong impact on the assessment of the shock geometry. In fact, looking at the PDF distribution of $\theta_{Bn}$ values obtained using the same averaging parameters described for the quiet event, we find a much larger spread with a more gaussian behaviour, with smaller values of skewness (~ $10^{-2}$) with respect to the quiet event of October 11. The average $\theta_{Bn}$ computed with the mixed mode method is consistent, and between 42 and 45°. These results are different than the ones found using magnetic coplanarity, confirming the strong variability of the parameter estimation for this event. However, it is possible to notice the fact that the systematic approach to upstream/downstream averages yields to an accurate estimation of the shock geometry while addressing the sensitivity to the averaging window choice, and therefore accuracy through the $\sigma_{\theta_{Bn}}$ parameter.

The two above examples represent two very recent shock observations, and their study is extremely interesting in their own respect, and will be part of a separate work. Furthermore, for other shock parameters estimated, such as Mach numbers and compression ratio, a systematic approach has been followed as well. The results of such approach are reported in the Appendix of this work (see Figure 7), and the software used to compute them is included in the Python package SerPyShock.

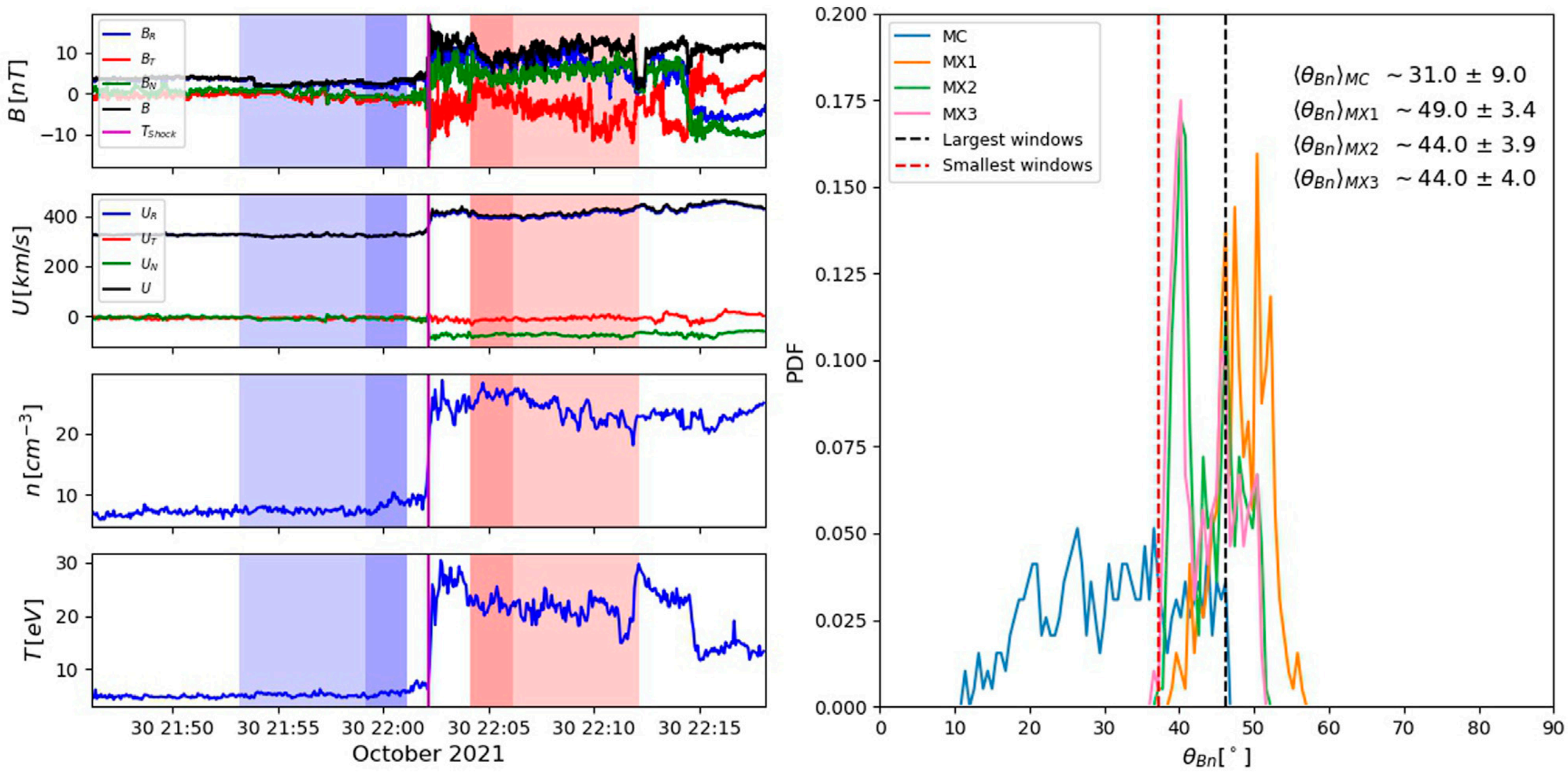

**FIGURE 6**
The structured event of 30 October 2021 seen by Solar Orbiter. The Figure is organised in the same fashion as Figure 5.

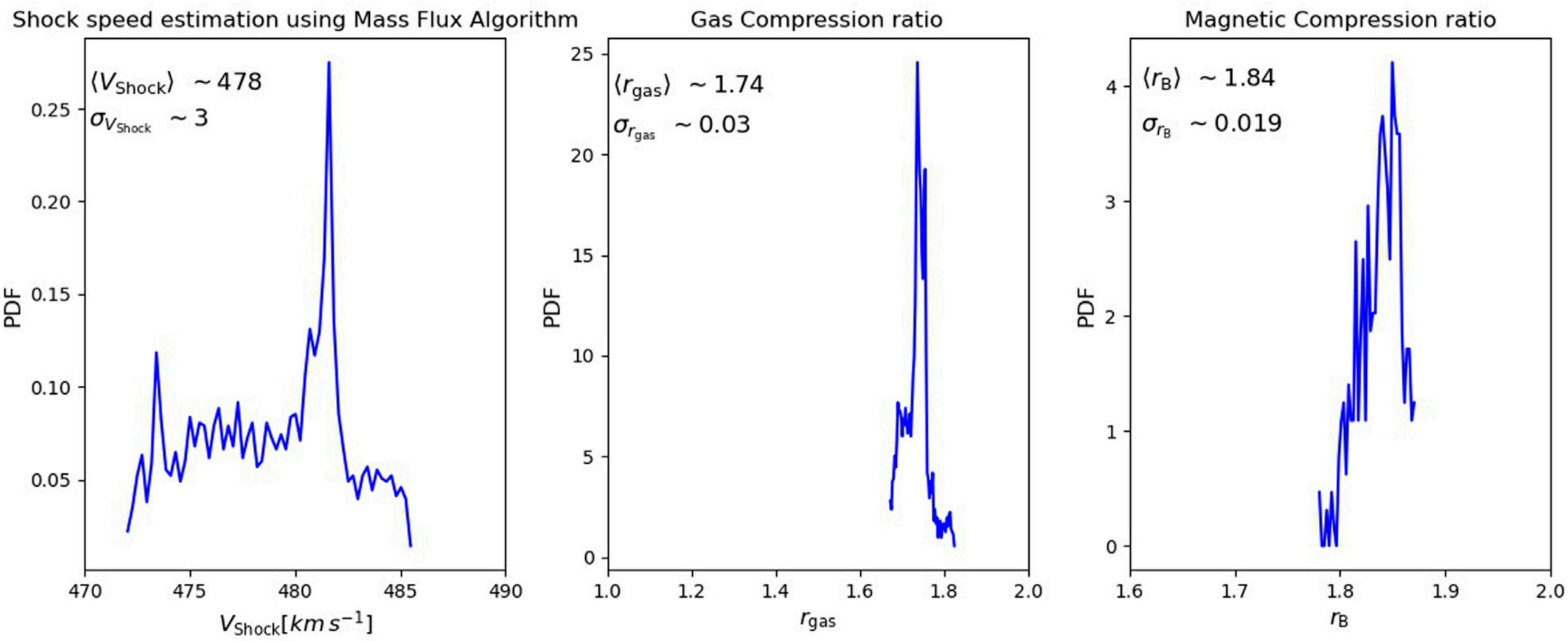

**FIGURE 7**
A test for shock speed, gas compression and magnetic compression ratio calculations using the systematic variation of averaging window technique for the quiet event of 11 October 2021 (see Figure 5).

## 3.4 MMS shock observations

The magnetospheric multiscale mission (MMS) provided the community with unprecedented multi-spacecraft high-resolution measurements of the magnetospheric environment (Burch et al., 2016). On 8 January 2018, while in the pristine solar wind, MMS observed a supercritical interplanetary shock. This interesting event was studied in detail by Cohen et al. (2019), reporting for the first time direct high temporal resolution near specularly reflected ions at an IP shock. During the shock crossing, the MMS spacecraft had small separation (~ 20 km), and it therefore represents a good opportunity to test our diagnostics on multi-spacecraft crossing of the same shock, not possible with Solar Orbiter. Figure 8 shows magnetic field measurement for the four different MMS spacecraft around the shock crossing time, and the analysis for the shock obliquity is carried out using magnetic coplanarity in the right hand-panel. Here, the range of upstream/downstream time windows has been set to be close to the averaging windows used by Cohen et al. (2019). We find strong agreement between the shock obliquity estimated at each shock crossing (Figure 8B), with a $\theta_{Bn}$ value consistent with the one reported in Table 1 of Cohen et al. (2019), that is of 67°.

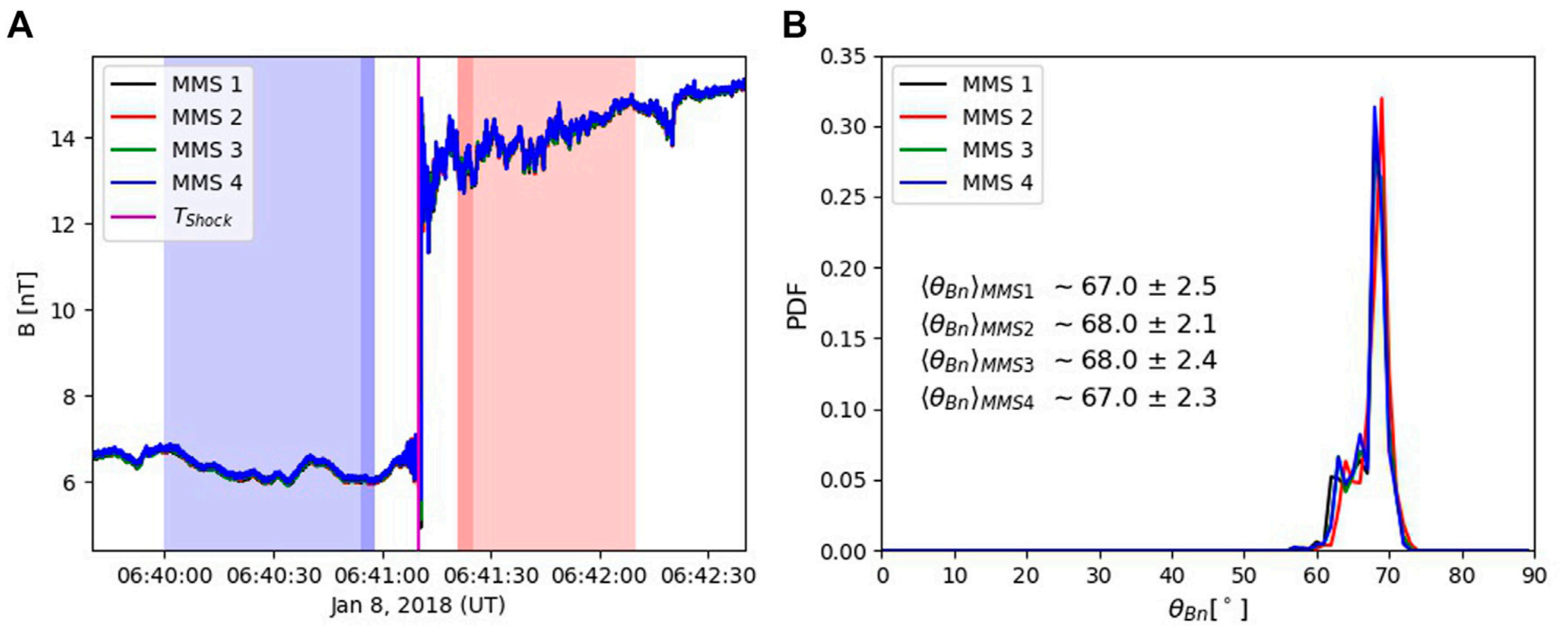

FIGURE 8
**(A)**: Magnetic field magnitude observed by the four MMS spacecraft of 8 January 2018. The purple line represents the time of the shock crossing. **(B)**: PDF distributions of $\theta_{Bn}$ values obtained with the magnetic coplanarity method for each spacecraft.

# 4 Conclusion

In this work, we revisited shock parameter estimation using single spacecraft signals. Starting from the early seminal works of Balogh et al. (1995), a method involving a systematic variation of upstream/downstream averaging windows has been implemented, yielding to an ensemble of shock parameters estimations for a shock crossing, as opposed to a single value that corresponds to a particular choice of upstream/downstream windows. With such a statistic of shock parameters, it is possible to address their mean value as a more accurate parameter estimation, and the standard deviation as a measure of uncertainty/sensitivity to the choice of averaging windows. We discussed the implication of adopting such an approach.

We started by introducing the main shock parameters estimation techniques tested throughout this work (Section 2), with particular emphasis on shock normal estimation methods, reviewing the theoretical framework for Magnetic Coplanarity (MC) and Mixed Mode (MX1-2-3) methods, closely following previous, more extensive documentations (Paschmann and Schwartz, 2000; Hietala, 2012; Burgess and Scholer, 2015). Throughout the work, we focus on shock geometry estimation (shock normals and consequently $\theta_{Bn}$ angles).

The ensemble technique to compute shock parameters is then introduced on the simplest possible test cases, namely synthetic timeseries of Rankine-Hugoniot compliant shock transitions (Section 3.1). Such synthetic shocks have been analysed both in the purely analytic case, and also adding white noise, to mimic some level of uncertainty in the fields across the shock transition. The technique is tested, and the method followed is highlighted. It is important to note that, due to the nature of the shock parameters estimation involved, that suppose an upstream/downstream averaging region as close as possible to the shock with an exclusion zone containing the shock itself, our systematic averaging variation is done in such a way that, after starting from the smallest possible upstream (downstream) windows, these are enlarged at every iteration by some extent. In this way, points close to the shock transition are included multiple times in the parameters statistics. Using random window generation upstream/downstream, it may be possible to accidentally choose windows that are very far from the shock transition, violating the premise of the parameter estimation techniques analysed here (Hietala, 2012).

An hybrid kinetic, self-consistent simulation of an oblique, supercritical shock is then presented as a further test for shock geometry evaluation. These simulations are very successful at reproducing many features of collisionless shock transitions, including shock front instabilities and the presence of self-consistently generated upstream/downstream fluctuations. The advantage is that a nominal $\theta_{Bn}$ is chosen for the shock as an initial condition for the simulation. Using a virtual satellite crossing the shock transition, we applied the single-spacecraft techniques mentioned above, a numerical experiment that has not been reported in previous literature, to the best of our knowledge. We found that at the early stage of the shock simulation, where upstream/downstream wave activity is low and the shock front is quasi-laminar, the PDF distribution of shock normal angles peaks very well around the nominal $\theta_{Bn}$. For later simulation stages, where the shock front is unstable and the upstream/downstream regions filled with fluctuations related to unstable particle distributions generated at the shock, the distribution of $\theta_{Bn}$ angles widens remarkably. Interestingly though, even with a single (virtual) spacecraft crossing of the shock front, the averaged $\theta_{Bn}$ values obtained from the distribution for MC and MX1-2-3 techniques are rather close to the nominal one posed as an initial condition. Such an experiment highlights, on one hand, the capabilities of the ensemble approach to shock parameter estimation, and on the other hand the fact that single spacecraft crossings, even though limited, contain extremely valuable information about shock features. This numerical experiment will be extended in future works, including crossings at different angles with respect to the shock normal, as well as employing simulations using full three-dimensional geometry.

In Section 3.3, we applied the systematic shock parameter estimation technique to two recent fast forward, CME-driven shocks observed by Solar Orbiter. These events are interesting in their own respect and will be object of detailed investigation addressing the particle behaviour around the shock transitions. Shock parameters were estimated for the two different shocks, one characterised by rather low Mach numbers and a quiet upstream/downstream environment, named here the October 11 “quiet event”, and another one characterised by rather high Mach numbers, that is propagating though a structured upstream and downstream medium. The technique performed well, revealing sharper PDF distributions for $\theta_{Bn}$ for the quiet event with respect to the broader PDFs observed for the structured event. These examples show that this systematic way of estimating shock parameters is consistent with the expected sensitivity to the averaging window choice, and represent a step forward with respect to choosing one single combination of upstream/downstream windows. Future works will compare the results obtained with this approach with multi-spacecraft techniques to evaluate parameters such as, for example, the shock speed.

We have developed an open-source Python software package, SerPyShock, that can be used to perform an analysis of shock wave properties similar to the one presented in this study. This package provides the code we developed for such shock analyses, an example script, and the data loaders needed to work with Solar Orbiter *in situ* datasets. The source code of the package is provided in a publicly available GitHub repository https://github.com/trottadom/SerPyShock and it is licensed under the GNU General Public License version 3. We plan to extend the capabilities of this software package by including other useful techniques for shock physics investigation, such as, for example, a set of routines looking at the properties of energetic particles.

## Data availability statement

The original contributions presented in the study are publicly available. The software package SerPyShock used to make all the Figures and analyses is available at https://github.com/trottadom/SerPyShock. The hybrid simulation datasets used in Section 3.2 and Figures 3, 4 are available to download here: https://doi.org/10.5281/zenodo.6913596. The Solar Orbiter data used are available for download at https://soar.esac.esa.int/soar/. The MMS data used in Figure 7 can be downloaded at https://cdaweb.gsfc.nasa.gov.

## Author contributions

DT and HH discussed and laid the foundations for the technique presented here. DT, HH, and TH decided the structure of this work and the analysis plan. DT performed the analyses and produced the figures. DT performed the hybrid kinetic simulation, looking at virtual spacecraft signals in collaboration with FV. EK and DT tested and expanded the diagnostics proposed in our work. The SerPyShock Python code has been developed by DT. LV and JG developed the Solar Orbiter data loaders included in the software. JG, AK, and DP provided support for the Python development. ND, FV, and RV helped to discuss the relevance of this work for the field. All authors contributed to the global interpretation of the results, as well as to their discussion and presentation in the manuscript. All authors revised the manuscript before submission.

## Funding

This work has received funding from the European Unions Horizon 2020 research and innovation pro-gramme under grant agreement No. 101004159 (SERPENTINE, www.serpentine-h2020.eu). Part of this work was performed using the DiRAC Data Intensive service at Leicester, operated by the University of Leicester IT Services, which forms part of the STFC DiRAC HPC Facility (www.dirac.ac.uk), under the project "dp031 Turbulence, Shocks and Dissipation in Space Plasmas". AK acknowledges financial support from NASA NNN06AA01C (SO-SIS Phase-E) contract.

## Acknowledgments

DT thanks Dr. A. Dimmock for the fruitful discussions around the capabilities of the toolbox. DT would like to thank Prof. D. Burgess for the support in performing the hybrid kinetic simulation. ND is grateful for support by the Turku Collegium for Science, Medicine and Technology of the University of Turku, Finland.

## Conflict of interest

The authors declare that the research was conducted in the absence of any commercial or financial relationships that could be construed as a potential conflict of interest.

## Publisher's note

## Appendix: Shock speed estimation and other parameters

This work demonstrates the use and implications of having a systematic variation of upstream/downstream averaging windows when addressing shock parameters using single spacecraft crossings. In the discussion, our main focus has been around the estimation of shock normal vectors and $\theta_{Bn}$ angles to address the shock geometry, crucial for many features of collisionless shocks. However, other important parameters can be estimated using the same ensemble approach for the averaging operation. In the SerPyShock software released together with this work, we also provide routines for the systematic computation of shock speed using the mass flux algorithm $V_{\mathrm{shock}}$, and the shock gas and magnetic compression ratios $\mathrm{r_{gas}}$ and $\mathrm{r_B}$. The definitions of these parameters are given in Section 2. An example of usage of such routines is shown in Figure 8, where they have been applied to the shock observed by Solar Orbiter on 11 October 2021 (i.e., the "quiet event" discussed in Section 3.3).